\documentclass[twocolumn,amsmath,amssymb,footinbib,pra,showpacs]{revtex4}

\usepackage{graphicx}
\usepackage{picture}
\usepackage{epstopdf}
\usepackage{pdfpages}
\usepackage{microtype}
\usepackage{tikz}
\usepackage{hyperref} 

\begin{document}

\title{Comparing models for the ground state energy of a trapped
  one-dimensional Fermi gas with a single impurity}
\date{\today}

\author{N.~J.~S. \surname{Loft}}
\affiliation{Department of Physics and Astronomy, Aarhus University, DK-8000 Aarhus C,  Denmark}
\author{L.~B. \surname{Kristensen}}
\affiliation{Department of Physics and Astronomy, Aarhus University, DK-8000 Aarhus C,  Denmark}
\author{A.~E. \surname{Thomsen}}
\affiliation{Department of Physics and Astronomy, Aarhus University, DK-8000 Aarhus C,  Denmark}
\author{N.~T. \surname{Zinner}}
\affiliation{Department of Physics and Astronomy, Aarhus University, DK-8000 Aarhus C,  Denmark}

\begin{abstract}
  We discuss the local density approximation approach to calculating
  the ground state energy of a one-dimensional Fermi gas containing a
  single impurity, and compare the results with exact numerical values
  that we have for up to 11 particles for general interaction
  strengths and up to 30 particles in the strongly interacting case.
  We also calculate the contact coefficient in the strongly
  interacting regime. The different theoretical predictions are
  compared to recent experimental results with few-atom
  systems. Firstly, we find that the local density approximation
  suffers from great ambiguity in the few-atom regime, yet it works
  surprisingly well for some models. Secondly, we find that the strong
  interaction theories quickly break down when the number of particles
  increase or the interaction strength decreases.
\end{abstract}
\date{\today}
\maketitle

\section{Introduction}
As cold atomic gas capabilities advance
\cite{bloch2008,lewenstein2007}, a large number of interesting quantum
systems can now be built using the cold atom toolbox
\cite{esslinger2010,baranov2012,zinner2013}. The structure and
dynamics of one-dimensional quantum systems is a specific venue in
which cold atoms have driven a new surge of excitement as a number of
important models of many-body quantum physics can now be probed
experimentally
\cite{olshanii1998,moritz2003,stoferle2004,kinoshita2004,paredes2004,kinoshita2006,haller2009,haller2010,pagano2014}.
This includes the famous strongly interacting Bose systems studied by
Tonks \cite{tonks1936} and Girardeau \cite{girardeau1960}.  More
recently, it has become possible to push these studies to smaller
particle numbers \cite{serwane2011} so that few-body physics in
one-dimensional traps with controllable interactions can now be
realized. A number of such studies in recent years using two-component
Fermi systems have probed such concepts as fermionization
\cite{zurn2012}, pairing \cite{zurn2013}, polarons \cite{wenz2013},
and the Hubbard \cite{murmann2015a} and Heisenberg models
\cite{murmann2015b}, all in the limit where the system contains just a
few particles. These developments have sparked a lot of interest in
the few-body community and a number of new theoretical studies looking
at various aspects of these two-component Fermi systems in one
dimension have emerged
\cite{guan2009,yang2009,girardeau2010,guan2010,rubeni2012,brouzos2013,bugnion2013,astrakharchik2013,gharashi2013,sowinski2013,gharashi2014,artem2014,lindgren2014,volosniev2014,deuret2014,cui2014,doggen2014,loft2015,unal2015,lundmark2015,gharashi2015,volosniev2015,yang2015,sowinski2015,levinsen2014,artem2015f,grining2015}.

In the present paper we are interested in the recent experiments that
have shown how one may build a Fermi system consisting of an impurity
and a Fermi sea one particle at a time \cite{wenz2013}. This amounts
to building a so-called Fermi polaron from a few-body perspective.
The experimental results presented in Ref.~\cite{wenz2013} suggest
that already for a system consisting of only six particles (five
identical fermions and a single impurity of the same mass as the other
particles) one can see the emergence of a many-body system taking
place. This is a rather remarkable result and in the present paper we
would like to explore this in more detail. The experimental results
consists of a series of measurements of the energy for different
interaction strengths and for different system sizes and have
previously been compared to theoretical predictions in the literature
\cite{astrakharchik2013}.  However, there are some ambiguities in this
comparison arising from the fact that the experimental system is in a
trap while a lot of theoretical work considers a homogeneous
system. Proposals for merging the homogeneous results with a trapped
system are known and one of our goals here is to assess how well they
work as a function of particle number and interaction strength. We
will take advantage of the fact that we have access to both very
accurate numerical calculations for all interaction strengths for up
to eleven particles and in the strongly interacting regime for up to
thirty particles in a harmonic trap. Using these results we will shed
some light on the questions concerning how many particles constitutes
`many' in one-dimensional impurity systems and when such systems can
be considered in the strongly interacting regime.

\section{Model}
We consider a system of $N$ spin-$\tfrac{1}{2}$ fermions with mass $m$
confined in a one-dimensional (1D) harmonic trap with oscillator
frequency $\omega$. The fermions in the gas are spin-polarized with
$N-1$ particles in, say, the spin-up state, while the remaining
particle (the impurity) is in the spin-down state. The particles
interact via a short-range interaction modelled by a delta function of
coupling strength $g$. The system is described by the Hamiltonian
\begin{equation}
  \label{eq:hamiltonian}
  H = \sum\limits_{i=1}^N
  \left[ - \frac{\hbar^2}{2m} \frac{\partial^2}{\partial x_i^2}
         + \frac{1}{2} m \omega^2 x_i^2 \right]
  + g \sum\limits_{i < j} \delta(x_i - x_j) \; .
\end{equation}
In the interaction term, the sum runs over all pairs of particles, but
due to the antisymmetry of the total wave function under exchange of
any two majority particles, the majority particles do not couple to
each other. Thus the interaction term above is equivalent to only
summing over the $N-1$ pairs consisting of the impurity and each of
the majority particles. As has been recently demonstrated
experimentally, the interaction strength may be varied from small to
large positive and negative values \cite{zurn2012}.  We will limit
ourselves to a repulsive interaction, $g > 0$.  Throughout this paper,
we use harmonic oscillator units where $\omega = m = \hbar = 1$.

\section{Strong coupling regime}
We start by discussing the Tonks-Girardeau (TG) limit in which the
interaction between the particles is infinitely strong, $1/g
\rightarrow 0$. In this limit, the ground state becomes
$N!/((N-1)!\cdot1!)  = N$-fold degenerate with the energy
$E_\infty^\text{t} = N^2/2$, and the wave function vanish whenever any
two particles meet. The superscript t denotes the fact that the system
is trapped, and the subscript $\infty$ refers to the interaction being
infinitely strong. Notice that the degenerate energy in the ground
state manifold is also the ground state energy of an ideal
non-interacting Fermi gas of $N$ particles in a harmonic trap,
i.e. the particles occupy the $N$ lowest-lying energy states in the
trap.

The degeneracy is lifted by moving slightly away from $1/g = 0$, and
the wave function is lifted slightly away from zero at the surfaces
where the impurity meets a majority particle. This allows the impurity
to switch position with its neighbors, and we can therefore think of
the particles as occupying sites on a lattice. The system can be
described by an effective Heisenberg spin chain model as discussed in
\cite{volosniev2014,deuret2014,volosniev2015,levinsen2014}.  To linear
order in $1/g \ll 1$, we may cast the Hamiltonian as
\begin{equation}
  \label{eq:heisenberg}
  H = E_\infty^\text{t} - \frac{1}{g}
  \sum\limits_{k=1}^{N-1} \frac{\alpha_k}{2}
  \left[ 1 - \boldsymbol{\sigma}^k \cdot \boldsymbol{\sigma}^{k+1} \right] \; ,
\end{equation}
where $\boldsymbol{\sigma}^k = (\sigma_x^k, \sigma_y^k, \sigma_z^k)$
are the Pauli matrices acting on the spin of the particle at site $k$,
and $\alpha_k$ is a geometric coefficient determined by the trap
potential. These are the same $\alpha$-coefficients discussed in
\cite{volosniev2015}, where the case of a general confining potential,
$V(x)$, and an arbitrary number of spin-up and spin-down particles is
considered. The general case captures the behavior of both bosonic and
fermionic particles near the TG limit. In Ref.~\cite{loft2016b} it is
shown that the geometric coefficient $\alpha_k$ generally can be
expressed as:
\begin{align}
  \label{eq:geometric-final}
  \alpha_k = \; &2
  \sum_{i=1}^N \sum_{j=1}^N \sum_{l=0}^{N-1-k}
  \frac{(-1)^{i+j+N-k}}{l!}
  {N-l-2 \choose k-1} \nonumber\\
  &\times \int_a^b \textrm{d}x \,
  \frac{2m}{\hbar^2} \big( V(x) - E_i \big) \, \psi_i (x) \,
  \frac{\textrm{d}\psi_j}{\textrm{d}x} \nonumber\\
  &\times
  \left[
    \frac{\partial^l}{\partial\lambda^l}
    \det \big[ (B(x) - \lambda \textbf{I})^{(ij)} \big]
  \right]_{\lambda = 0} \nonumber\\
  &+ \sum_{i=1}^N \left[ \frac{\textrm{d}\psi_i}{\textrm{d}x}
  \right]^2_{x=b}\; .
\end{align}
To compute this we need access to the $N$ lowest-energy
single-particle wavefunctions $\psi_i(x)$ solving the Schr\"odinger
equation $\big[\tfrac{-\hbar^2}{2m} \tfrac{\partial^2}{\partial x^2} +
V(x)\big]\psi_i = E_i \psi_i$. In the present case, this is the
analytically solved Schr\"odinger equation for a single particle in a
harmonic oscillator. In Eq.~\eqref{eq:geometric-final} $B(x)$ is a $N
\times N$ symmetric matrix with the $mn$'th entry given by $\int_a^x
\textrm{d}y \, \psi_m(y) \psi_n(y)$, and $(\;)^{(ij)}$ defines a minor
obtained by removing the $i$'th column and the $j$'th row. For most
potentials such as smooth potentials like the present case, the
integration limits will be $b = \infty = - a$, and for some
non-analytic potentials such as a hard box, the integration will be
confined to some finite region of space. For symmetric confining
potentials, as the one studied in this paper, the coefficients are
also symmetric, $\alpha_k = \alpha_{N-k}$.

The main motivation for the present work is that we are able to
determine these geometric $\alpha$-coefficients exactly in the general
case of an arbitrary potential, see Ref.~\cite{loft2016b} for a
detailed presentation of our method including a derivation of
Eq.~\eqref{eq:geometric-final}. The method is implemented as a highly
efficient numerical program released as open source
software~\footnote{The program and source code can be downloaded
  \href{http://phys.au.dk/forskning/forskningsomraader/amo/few-body-physics-in-a-many-body-world/conan/}{here}.}. We
stress the importance of our capability to perform these calculations
numerically exact; we do not use Monte Carlo integration or any other
approximate method. For this paper, we calculate the geometric
coefficients numerically exact for a harmonic potential for $N \leq
30$. The coefficients are given in Section~\ref{appa}.Calculating the
coefficients for such a high number of particles as 30 is an extremely
demanding computational task. 
However, the method we have developed enables us to perform
these calculations on a small laptop to a very high level of precision
in a matter of hours for 30 particles. For a few particles, the
calculations are done almost immediately. 

Since the harmonic potential is such a well-studied confining
potential, this allows us to compare our exact results with the
approximate methods studied in the literature. In particular, we are
interested in comparing methods for calculating the ground state
energy in the vicinity of the TG limit. To linear order in $1/g$, we
write the ground state energy as
\begin{equation}
  \label{eq:gs-energy}
  E_g^\text{t} = E_\infty^\text{t} - \frac{1}{g} \mathcal{C} \; ,
\end{equation}
where $\mathcal{C}$ is the eigenvalue of the sum in
Eq.~\eqref{eq:heisenberg} for the ground state, i.e. the largest
eigenvalue. We denote $\mathcal{C}$ the contact coefficient
\cite{tan2008,valiente2012}. Access to the exact geometric
coefficients $\alpha_i$ allows us to diagonalize the Hamiltonian in
Eq.~\eqref{eq:heisenberg} and calculate the contact coefficient
exactly. For $N \leq 30$ we compute $\mathcal{C}$ and want to compare
our results with previous approximate formulas for this constant.

\subsection{Previous results}
Now we will review some previous calculations of the contact
coefficient $\mathcal{C}$ in Eq.~\eqref{eq:gs-energy}. However, since
the previous theories may not take the trap into account, we will have
to adapt the existing formulas in order to directly compare them with
our results. In the following, we will distinguish between the
\emph{free} system and the (harmonically) \emph{trapped}
system. Furthermore, we will discuss different regimes of the
interaction strength $g$ and the particle number $N$.

A pioneering piece of work was done by McGuire \cite{mcguire1965} who
calculated the ground state energy for a free 1D Fermi gas with a
single impurity for an arbitrary interaction strength $g$ subject to
periodic boundary conditions. Thus the trap needs to be taken into
account. This problem was recently addressed by Astrakharchik and
Brouzos who applied the local density approximation (LDA) to McGuire's
result in order to obtain an expression for a harmonically trapped
system \cite{astrakharchik2013}. However, McGuire assumes that the
number of particles $N$ is even and large. This subtlety does not
appear to be explicitly discussed in Ref.~\cite{astrakharchik2013}.
Namely, the implicit condition that $N$ is very large makes it
interesting to compare the result with our exact results for small
$N$. To get the desired contact coefficient $\mathcal{C}$, we should
expand Astrakharchik and Brouzos' LDA result around $1/g = 0$ to first
order and read off the coefficient.

McGuire's energy expression was found for a free interacting system
subject to periodic boundary conditions (PBC). It would be interesting
to consider a hard wall boundary condition (HWC), too. This problem
was solved by Oelkers {\it et al.} for a free gas in the strong
coupling regime \cite{oelkers2006} which is exactly the regime we are
interested in. In fact Oelkers {\it et al.} found results for both PBC
and HWC for finite (even) $N$, and not just in the limit of large
$N$. Just as Astrakharchik and Brouzos applied LDA to McGuire's free
result, we shall apply LDA to Oelkers' free results. Since the LDA
method maps the free system to a trapped system, we would intuitively
think that HWC would produce a better final result, since a hard wall
would resemble a harmonic potential more that a periodic
boundary. This will turn out not to be the case, but as we shall also
see, these LDA results are subject to high uncertainties in the small
$N$ regime. Nevertheless this is an important regime as current
experiments study the behaviour of few particles.

The starting point for our analysis are the results summarized in
Table~\ref{tab:summary}. Here we have recorded whether the energy is
for a harmonically trapped or a free system, which boundary condition
is used and which values of the interaction strength and particle
numbers it is valid for.

\begin{table}[htbp]
  \centering
  \begin{tabular}{l @{\hspace{12pt}} c @{\hspace{12pt}} c
      @{\hspace{12pt}} c @{\hspace{12pt}} c}
    Reference & Trapped & BC & $g$ & $N$ \\
    \hline
    McGuire \cite{mcguire1965}       & No  & PBC & any & $N \rightarrow \infty$  \\
    Astra. \cite{astrakharchik2013}  & Yes & PBC & any & $N \rightarrow \infty$  \\
    Oelkers \cite{oelkers2006}       & No  & PBC & $g \rightarrow \infty$ & even \\
    Oelkers \cite{oelkers2006}       & No  & HWC & $g \rightarrow \infty$ & even
  \end{tabular}
  \caption{Summary of some previous results.}
  \label{tab:summary}
\end{table}

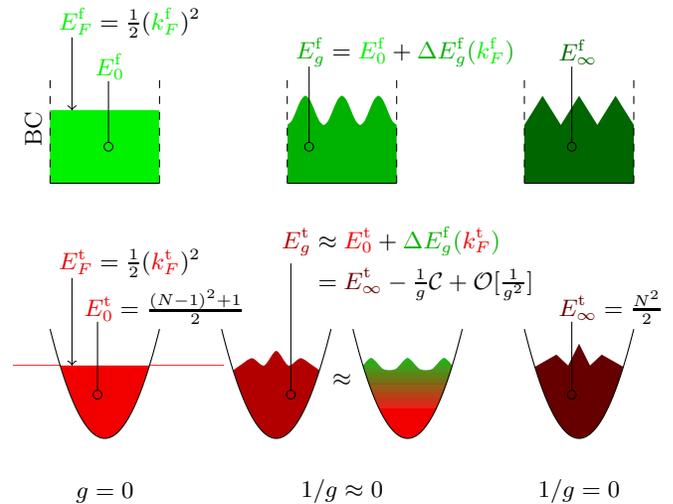
\begin{figure}[tbp]
  \centering
  \hspace*{-1.2mm}
  \begin{tikzpicture}[scale=.97]

    \clip (-.5,0) rectangle (8.5,8);

    \begin{scope}[shift={(0,4.5)}]; 
      \fill [black!5!green] (0,0) rectangle (1.5,1);
      \draw (0,0) -- (1.5,0);
      \draw [dashed] (0,0) -- (0,1.5);
      \draw [dashed] (1.5,0) -- (1.5,1.5);
      \node [above, rotate=90] at (0,.75) {BC};
      \node [right] at (0,2.2)
      {$\textcolor{black!5!green}{E_F^\text{f}} = \tfrac{1}{2}(
        \textcolor{black!5!green}{k_F^\text{f}})^2$};
      \draw [->] (.3,2) -- (.3,1);
      \node [right] at (.5,1.6)
      {$\textcolor{black!5!green}{E_0^\text{f}}$};
      \draw (.8,.5) circle (.06);
      \draw [-] (.8,1.4) -- (.8,.56);
    \end{scope}

    \begin{scope}[shift={(3.25,4.5)}]; 
      \begin{scope}
        \clip (0,0) -- plot[smooth,domain=0:1.5](\x,{1+.2*cos(deg(pi+\x*12/3*pi))})
        -- (1.5,0);
        \fill [black!30!green] (0,0) rectangle (2,2);
      \end{scope}
      \draw (0,0) -- (1.5,0);
      \draw [dashed] (0,0) -- (0,1.5);
      \draw [dashed] (1.5,0) -- (1.5,1.5);
      \node [right] at (0,1.8)
      {$\textcolor{black!30!green}{E_g^\text{f}} =
        \textcolor{black!5!green}{E_0^\text{f}} +
        \textcolor{black!30!green}{\Delta E_g^\text{f}(}
        \textcolor{black!5!green}{k_F^\text{f}}
        \textcolor{black!30!green}{)}$};
      \draw (.3,.5) circle (.06);      
      \draw [-] (.3,1.6) -- (.3,.56);
    \end{scope}

    \begin{scope}[shift={(6.5,4.5)}]; 
      \begin{scope}
        \clip (0,0) -- (0,.8) -- (1/6*3/2,1.2) -- (2/6*3/2,.8) --
        (3/6*3/2,1.2) -- (4/6*3/2,.8) -- (5/6*3/2,1.2) -- (6/6*3/2,.8)
        -- (1.5,0); \fill [black!60!green] (0,0) rectangle (2,2);
      \end{scope}
      \draw (0,0) -- (1.5,0);
      \draw [dashed] (0,0) -- (0,1.5);
      \draw [dashed] (1.5,0) -- (1.5,1.5);
      \node [right] at (.35,1.8)
      {$\textcolor{black!60!green}{E_\infty^\text{f}}$};
      \draw (.65,.5) circle (.06);      
      \draw [-] (.65,1.6) -- (.65,.56);
    \end{scope}

    \begin{scope}[shift={(.75,1)}]; 
      \begin{scope}
        \clip (-1.633,1) -- plot[smooth,domain=-1.633:1.633](\x,{2.6667*\x*\x})
        -- (1.633,1);
        \fill [black!5!red] (-2,0) rectangle (2,1);
      \end{scope}
      \draw plot[smooth,domain=-.75:.75](\x,{2.6667*\x*\x});
      \node [right] at (-.75,2.4)
      {$\textcolor{black!5!red}{E_F^\text{t}} = \tfrac{1}{2}(
        \textcolor{black!5!red}{k_F^\text{t}})^2$};
      \draw [->] (.3-.75,2.2) -- (.3-.75,1);
      \node [right] at (-.3-.1,1.8)
      {$\textcolor{black!5!red}{E_0^\text{t}} = \tfrac{(N-1)^2+1}{2}$};
      \draw (-.1,.6) circle (.06);
      \draw [-] (-.1,1.6) -- (-.1,.66);
    \end{scope}

    \begin{scope}[shift={(3.1,1)}]; 
      \begin{scope}
        \clip (-2,2) -- plot[smooth,domain=-2:2](\x,{2.6667*\x*\x})
        -- (2,2);
        \clip (-2,0) -- plot[smooth,domain=-2:2](\x,{1+.2*cos(deg(\x*3*pi))*cos(deg(\x*2*pi))}) -- (2,0);
        \fill [black!30!red] (-2,0) rectangle (2,1.5);
      \end{scope}
      \draw plot[smooth,domain=-.75:.75](\x,{2.6667*\x*\x});
      \node [right] at (-.4+.3,2.4)
      {$\begin{aligned}
        \textcolor{black!30!red}{ E_g^\text{t}}
        &\approx
        \textcolor{black!5!red}{E_0^\text{t}} +
        \textcolor{black!30!green}{\Delta E_g^\text{f}(}
        \textcolor{black!5!red}{k_F^\text{t}}
        \textcolor{black!30!green}{)} \\
        &=
        \textcolor{black!60!red}{E_\infty^\text{t}} -
        \tfrac{1}{g} \mathcal{C} + \mathcal{O}[\tfrac{1}{g^2}]
      \end{aligned}$};
      \draw (-.1+.3,.6) circle (.06);      
      \draw [-] (-.1+.3,2.4) -- (-.1+.3,.66);
    \end{scope}

    \begin{scope}[shift={(4.9,1)}]; 
      \begin{scope}
        \clip (-2,2) -- plot[smooth,domain=-2:2](\x,{2.6667*\x*\x}) --
        (2,2);
        \clip (-2,0) -- plot[smooth,domain=-2:2](\x,{1+.1*cos(deg(\x*4*pi))}) --
        (2,0);
        \fill [black!5!red] (-2,0) rectangle (2,.401);
        \shade [top color=black!30!green, bottom color=black!5!red]
        (-2,.4) rectangle (2,1.1);
      \end{scope}
      \draw plot[smooth,domain=-.75:.75](\x,{2.6667*\x*\x});
    \end{scope}
    
    \begin{scope}[shift={(7.25,1)}]; 
      \begin{scope}
        \clip (-2,2) -- plot[smooth,domain=-2:2](\x,{2.6667*\x*\x})
        -- (2,2);
        \clip (-.75,0) -- (-.75,.9) -- (-.31,1.15) -- (-.15,1) --
        (0,1.3) -- (.15,1) -- (.31,1.15) -- (.75,.9) -- (.75,0);
        \fill [black!60!red] (-2,0) rectangle (2,1.5);
      \end{scope}
      \draw plot[smooth,domain=-.75:.75](\x,{2.6667*\x*\x});
      \node [right] at (-.4,1.8)
      {$\textcolor{black!60!red}{E_\infty^\text{t}} = \tfrac{N^2}{2}$};
      \draw (-.1,.6) circle (.06);      
      \draw [-] (-.1,1.6) -- (-.1,.66);
    \end{scope}

    \node [above] at (4,1.6) {$\approx$};
    \node [above] at (.75,0) {$g=0$};
    \node [above] at (4,0) {$1/g \approx 0$};
    \node [above] at (7.25,0) {$1/g = 0$};
    
  \end{tikzpicture}
  \caption{Illustration of the local density approximation. The top
    part of the picture shows the free system solved for some boundary
    condition (green), whereas the bottom part shows the trapped
    system (red). As a way of representing an interacting sea of
    particles, we have drawn waves on the surface with increasing
    sharpness for increasing interaction strength. This should only be
    understood as an intuitive guide to the eye. The Fermi level is
    defined in the non-interacting case and denoted by an arrow, while
    the total energy of the system is denoted by a circle.}
  \label{fig:lda}
\end{figure}

\begin{figure*}[tbp]
  \centering
  \begin{tikzpicture}
    \shade[top color=white, bottom color=blue, opacity=.3] (.5,2.5)
    rectangle (2.5,3);
    \fill[color=blue, opacity=.3] (.5,.5) rectangle (2.5,2.5);
    \draw [->] (2.5,3.2) -- (2.5,.5) -- (.5,.5) -- (.5,3.5);
    \node [right] at (2.5,.5) {$i=0$};
    \draw (.5,1) -- (2.5,1);
    \node [right] at (2.5,1) {$i=\pm 1$};
    \draw (.5,1.5) -- (2.5,1.5);
    \node [right] at (2.5,1.5) {$i=\pm 2$};
    \draw (.5,2.5) -- (2.5,2.5);
    \node [right] at (2.5,2.5) {$i=\pm \tfrac{N-2}{2}$};
    \draw (.5,3) -- (2.5,3);
    \node [right] at (2.5,3) {$i=\pm \tfrac{N}{2}$};
    \draw [fill=cyan] (1.5,.5) circle (0.12);
    \draw [fill=cyan] (1.1,1) circle (0.12);
    \draw [fill=cyan] (1.9,1) circle (0.12);
    \draw [fill=cyan] (1.1,1.5) circle (0.12);
    \draw [fill=cyan] (1.9,1.5) circle (0.12);
    \draw [fill=cyan] (1.1,2.5) circle (0.12);
    \draw [fill=cyan] (1.9,2.5) circle (0.12);
    \draw [fill=white] (1.1,3) circle (0.12);
    \draw [fill=white] (1.9,3) circle (0.12);
    \fill [black] (1.5,1.8) circle (.03);
    \fill [black] (1.5,2) circle (.03);
    \fill [black] (1.5,2.2) circle (.03);
    \draw [line width = 2pt, cyan] (.5,2.5) -- (.5,3);
    \node [right] at (.2,3.8) {$E_i = \tfrac{1}{2}
      (k_i^\text{PBC})^2$};
    \node [left] at (.5,2.75) {$E_F^\text{f}$};
    \node [right] at (.4,0) {PBC, $N$ even};

    \fill[color=blue, opacity=.3] (5,.5) rectangle (7,2.5);
    \draw [->] (7,3.2) -- (7,.5) -- (5,.5) -- (5,3.5);
    \node [right] at (7,.5) {$i=0$};
    \draw (5,1) -- (7,1);
    \node [right] at (7,1) {$i=\pm 1$};
    \draw (5,2) -- (7,2);
    \node [right] at (7,2) {$i=\pm \tfrac{N-3}{2}$};
    \draw (5,2.5) -- (7,2.5);
    \node [right] at (7,2.5) {$i=\pm \tfrac{N-1}{2}$};
    \draw (5,3) -- (7,3);
    \node [right] at (7,3) {$i=\pm \tfrac{N+1}{2}$};
    \draw [fill=cyan] (6,.5) circle (0.12);
    \draw [fill=cyan] (5.6,1) circle (0.12);
    \draw [fill=cyan] (6.4,1) circle (0.12);
    \draw [fill=cyan] (5.6,2) circle (0.12);
    \draw [fill=cyan] (6.4,2) circle (0.12);
    \draw [fill=cyan] (5.6,2.5) circle (0.12);
    \draw [fill=white] (6.4,2.5) circle (0.12);
    \draw [fill=white] (5.6,3) circle (0.12);
    \draw [fill=white] (6.4,3) circle (0.12);
    \fill [black] (6,1.3) circle (.03);
    \fill [black] (6,1.5) circle (.03);
    \fill [black] (6,1.7) circle (.03);
    \node [right] at (4.7,3.8) {$E_i = \tfrac{1}{2}
      (k_i^\text{PBC})^2$};
    \draw [line width = 2pt, cyan] (5,2.46) -- (5,2.54);
    \node [left] at (5,2.5) {$E_F^\text{f}$};
    \node [right] at (5,0) {PBC, $N$ odd};

    \shade[top color=white, bottom color=blue, opacity=.3] (9.5,2.5)
    rectangle (11.5,3);
    \fill[color=blue, opacity=.3] (9.5,.5) rectangle (11.5,2.5);
    \draw [->] (11.5,3.2) -- (11.5,.5) -- (9.5,.5) -- (9.5,3.5);
    \draw (9.5,1) -- (11.5,1);
    \node [right] at (11.5,1) {$i= 1$};
    \draw (9.5,1.5) -- (11.5,1.5);
    \node [right] at (11.5,1.5) {$i= 2$};
    \draw (9.5,2.5) -- (11.5,2.5);
    \node [right] at (11.5,2.5) {$i=N-1$};
    \draw (9.5,3) -- (11.5,3);
    \node [right] at (11.5,3) {$i=N$};
    \draw [fill=cyan] (10.5,1) circle (0.12);
    \draw [fill=cyan] (10.5,1.5) circle (0.12);
    \draw [fill=cyan] (10.5,2.5) circle (0.12);
    \draw [fill=white] (10.5,3) circle (0.12);
    \fill [black] (10.5,2.2) circle (.03);
    \fill [black] (10.5,2) circle (.03);
    \fill [black] (10.5,1.8) circle (.03);
    \draw [line width = 2pt] (9.5,2.5) -- (9.5,3);
    \node [right] at (9.2,3.8) {$E_i = \tfrac{1}{2}
      (k_i^\text{HWC})^2$};
    \draw [line width = 2pt, cyan] (9.5,2.5) -- (9.5,3);
    \node [left] at (9.5,2.75) {$E_F^\text{f}$};
    \node [right] at (10,0) {HWC};

    \draw [->] (14,.5) -- (14,3.5);
    \begin{scope}[shift={(15.2,.5)}];

      \begin{scope}
        \clip (-.836660,1.75) -- plot[smooth,domain=-.836660:0](\x,{-2.5*\x^2}) -- (0,1.75);
        \fill[blue,opacity=.3] (-1,0) rectangle (1,2);
      \end{scope}
      \begin{scope}
        \clip (0,1.75) -- plot[smooth,domain=0:.836660](\x,{2.5*\x^2}) -- (.836660,1.75);
        \fill[blue,opacity=.3] (-1,0) rectangle (1,2);
      \end{scope}
      \begin{scope}
        \clip (-.948683,2.25) -- plot[smooth,domain=-.948683:-.836660](\x,{-2.5*\x^2})
        -- (-.836660,1.75) -- (0,1.75) -- (0,2.25);
        \shade[top color = white, bottom color=blue,opacity=.3] (-1,1.75) rectangle (1,2.25);
      \end{scope}
      \begin{scope}
        \clip (0,1.75) -- (.836660,1.75) -- plot[smooth,domain=.836660:.948683](\x,{2.5*\x^2})
        -- (.948683,2.25) -- (0,2.25);
        \shade[top color = white, bottom color=blue,opacity=.3] (-1,1.75) rectangle (1,2.25);
      \end{scope}
      
      \draw [domain=-1:0] plot (\x, {-2.5*\x^2});
      \draw [domain=0:1] plot (\x, {2.5*\x^2});
      \draw (-.316228,.25) -- (.316228,.25);
      \draw (-.547723,.75) -- (.547723,.75);
      \draw (-.836660,1.75) -- (.836660,1.75);
      \draw (-.948683,2.25) -- (.948683,2.25);
      \draw [fill=cyan] (0,.25) circle (0.12);
      \draw [fill=cyan] (0,.75) circle (0.12);
      \draw [fill=cyan] (0,1.75) circle (0.12);
      \draw [fill=white] (0,2.25) circle (0.12);
      \fill [black] (0,1.45) circle (.03);
      \fill [black] (0,1.25) circle (.03);
      \fill [black] (0,1.05) circle (.03);
      \node [right] at (.316228,.25) {$i=0$};
      \node [right] at (.547723,.75) {$i=1$};
      \node [right] at (.836660,1.75) {$i=N-2$};
      \node [right] at (.948683,2.25) {$i=N-1$};
    \end{scope}

    \node [right] at (13.7,3.8) {$E_i = \tfrac{1}{2}
      (k_i^\text{HO})^2$};
    \draw [line width = 2pt, cyan] (14,2.25) -- (14,2.75);
    \node [left] at (14,2.5) {$E_F^\text{t}$};
    \node [right] at (14.9,0) {HO};
  \end{tikzpicture}
  \caption{The Fermi level, $E_F^\text{f,t}$, should lie between the
    highest energy of the occupied states and the lowest energy of the
    unoccupied states for the non-interacting system. In this limit
    the $N-1$ states with lowest energy are filled up. Notice that in
    PBC for $N$ odd, the highest occupied state and the lowest
    unoccupied state share the same energy, and thus the Fermi level
    is well-defined, but that we have a certain freedom in the other
    cases. This freedoms leads us to define the tuning parameters
    $\lambda^\text{PBC}, \lambda^\text{HWC}, \lambda^\text{HO} \in
    [0;1]$.}
  \label{fig:states}
\end{figure*}
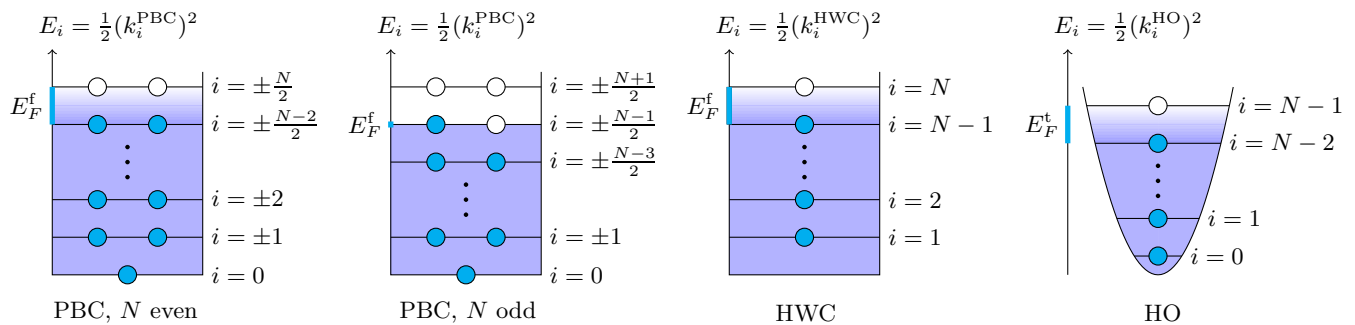

\subsection{The local density approximation}
The strategy to map a result from the free case to the trapped case is
by applying LDA. Therefore, we will briefly discuss LDA, and most
importantly we will encounter a weakness in the method for small
values of $N$.

Denote by $E_0^\text{f}$ ($E_0^\text{t}$) the energy of the
\emph{free} (\emph{trapped}), \emph{non-interacting} system. Suppose
that we know the energy states of these two systems, and that we
therefore can associate with them Fermi levels and Fermi momenta,
denoted $E_F^\text{f}$ ($E_F^\text{t}$) and $k_F^\text{f}$
($k_F^\text{t}$), respectively. Suppose furthermore that we know an
expression for the energy of the \emph{interacting} system, but only
in the \emph{free} case. Denote this energy by
\begin{equation}
  \label{eq:free-energy}
  E^\text{f}_g = E^\text{f}_0 + \Delta E_g^\text{f}(k_F^\text{f}) \; ,
\end{equation}
where $\Delta E_g^\text{f}$ is the correction due to interaction,
which will generally depend on $k_F^\text{f}$. Now, the quantity
sought after is the energy for the \emph{interacting} and
\emph{trapped} system. Within LDA, this is found by mapping
\begin{equation}
  \label{eq:lda-map-g0}
  E_0^\text{f} \mapsto E_0^\text{t}
  \qquad \text{and} \qquad
  k_F^\text{f} \mapsto k_F^\text{t} \; .
\end{equation}
The interaction correction in the trapped case retain its functional
form from the free case, but is now evaluated at the trapped Fermi
momentum instead of the free Fermi momentum. Thus the LDA expression
for the \emph{interacting} and \emph{trapped} system is
\begin{equation}
  \label{eq:lda-energy}
  E_ g^\text{t} \approx E^\text{t}_0 + \Delta E_g^\text{f}(k_F^\text{t}) \; .
\end{equation}
An illustrative sketch of the LDA procedure is shown on
Figure~\ref{fig:lda}.

We have just described the LDA method if our starting point was the
known non-interacting limit for both the free and trapped system. If,
however, we approached a finite interaction starting from the
infinitely strong interacting case, $1/g=0$, we would instead have
mapped
\begin{equation}
  \label{eq:lda-map-ginf}
  E_\infty^\text{f} \mapsto E_\infty^\text{t}
  \qquad \text{and} \qquad
  k_F^\text{f} \mapsto k_F^\text{t} \; ,
\end{equation}
where $E_\infty^\text{f}$ ($E_\infty^\text{t}$) is the energy of the
free (trapped) infinitely strongly interacting system, both assumed to
be known. We use this approach in Subsection~\ref{subsec:oelkers}.

The employment of LDA relies on the knowledge of the Fermi momenta,
$k_F^\text{f,t}$, for the free and trapped (non-interacting) systems,
or equivalently the Fermi levels, $E_F^\text{f,t}$. Now the problem is
that the Fermi level (also called the chemical potential) is not
well-defined if the systems contains a finite number of particles. We
know that the Fermi level should lie between the highest energy of the
occupied states and the lowest energy of the unoccupied states. But
exactly where between these two energies is irrelevant for the
occupancy of the states. The problem is illustrated on
Figure~\ref{fig:states} for the four relevant situations. In the
non-interacting limit, the $N-1$ identical fermions fill up the $N-1$
states of lowest energy.\footnote{The impurity is occupying the state
  of lowest energy, but this is irrelevant for the determination of
  the Fermi level.}  Recall that the momentum for a particle subject
to PBC, HWC or in a harmonic oscillator is given as
\begin{align*}
  &k_i^\text{PBC} = \frac{2\pi}{L} i \; ,
  &&i = 0, \pm 1, \pm 2, \dots \\[1mm]
  &k_i^\text{HWC} = \frac{\pi}{L} i \; ,
  &&i = 1, 2, 3, \dots \\[1mm]
  &k_i^\text{HO} = b^{-1}\sqrt{2i + 1} \; ,
  &&i = 0, 1, 2, \dots
\end{align*}
where $L$ is the length of the system and $b = \sqrt{\hbar/m\omega}$
is the standard oscillator length, $b=1$ in our units. Then the energy
at the Fermi levels can be described using the following expressions
for the Fermi momenta:
\begin{align*}
  &k_F^\text{PBC} =
  \begin{cases}
  \frac{\pi}{L} (N-2 +2\lambda^\text{PBC}) & \text{if $N$ is even} \\
  \frac{\pi}{L} (N-1) & \text{if $N$ is odd}
  \end{cases} \; , \\[1mm]
  &k_F^\text{HWC} = \frac{\pi}{L} (N-1 + \lambda^\text{HWC}) \; , \\[1mm]
  &k_F^\text{HO} = b^{-1}\sqrt{2(N - \tfrac{3}{2} + \lambda^\text{HO})} \; .
\end{align*}
Here $\lambda^\alpha \in [0,1]$ with $\alpha =$ PBC, HWC, HO is some
tuning parameter that allows us to probe the energies between that of
the highest occupied state and the lowest unoccupied state. Picking
$\lambda^\alpha = 0$ corresponds to picking the Fermi level at the
highest occupied state and $\lambda^\alpha = 1$ corresponds to taking
the Fermi level at the lowest unoccupied state. Another appealing
choice is $\lambda^\alpha = 1/2$, corresponding to taking the Fermi
level right in the middle.  For the time being, we will carry around
the tuning parameters, but at some point we would like to pick
specific values.

Note in particular that the ambiguity of the Fermi momentum disappears
in the thermodynamic limit, because the difference between the highest
occupied state and lowest unoccupied state becomes insignificant when
$N \rightarrow \infty$. This is reflected in the fact that the
ambiguity of the Fermi momentum is not discussed in
Ref.~\cite{astrakharchik2013}, where also the concepts of the Fermi
level and Fermi energy, which only equal each other in the
thermodynamic limit, are used somewhat interchangeably. But if we want
to derive energy expressions that apply to finite $N$, we should be
careful when choosing the Fermi momentum.

\subsection{McGuire, Astrakharchik and Brouzos (PBC, thermodynamic limit)}
In Ref.~\cite{mcguire1965} McGuire finds the following expression for
the ground state energy of the interacting, free system subject to
PBC:
\begin{equation}
  \label{eq:mg-deltae}
  \Delta E^\text{f}_g
  = \frac{(k_F^\text{PBC})^2}{2} \frac{\gamma}{\pi^2} \left[
    1 - \frac{\gamma}{4} +
    \left( \frac{\gamma}{2\pi} + \frac{2\pi}{\gamma} \right)
    \tan^{-1} \frac{\gamma}{2\pi} \right] \; ,
\end{equation}
where $\gamma = g \pi /k_F^\text{PBC}$. The result applies to all
values of the interaction strength $g$, but in deriving
Eq. \eqref{eq:mg-deltae} McGuire converts a sum to an integral letting
$N \rightarrow \infty$ and $L \rightarrow \infty$ with the density
$N/L$ held constant. This sum to integral conversion can be done in
several ways introducing some degree of freedom in $k_F^\text{PBC}$
consistent with the discussion in the previous section. Notice that
picking $\lambda^\text{PBC} = 1/2$ would set $k_F^\text{PBC} =
(N-1)\pi/L$ for all $N$. This would imply a vanishing interaction
correction for $N \rightarrow 1$ as it should.\footnote{We have to
  consider the limit of Eq.~\eqref{eq:mg-deltae} to avoid zero
  division at $k_F^\text{PBC} = 0$.}

We now sketch how Astrakharchik and Brouzos implement LDA on McGuire's
free energy  expression \eqref{eq:mg-deltae} to find the energy of the
trapped system \cite{astrakharchik2013}.\footnote{Notice that in
  Ref.~\cite{astrakharchik2013} the number of majority particles is
  denoted by $N$, and not $N-1$ as in our convention.} The LDA
expression for the energy of the trapped system \eqref{eq:lda-energy}
yields
\begin{align}
  \label{eq:astrakharchik}
  E_ g^\text{t} &\approx E^\text{t}_0 + \Delta E_g^\text{f}(k_F^\text{HO}) \nonumber\\
  &= \frac{(N-1)^2 +1}{2} \nonumber\\
  &+ \frac{(k_F^\text{HO})^2}{2} \frac{\gamma^\text{t}}{\pi^2} \left[
    1 - \frac{\gamma^\text{t}}{4} +
    \left( \frac{\gamma^\text{t}}{2\pi} + \frac{2\pi}{\gamma^\text{t}} \right)
    \tan^{-1} \frac{\gamma^\text{t}}{2\pi} \right] \; ,
\end{align}
with $\gamma^\text{t} = g\pi/k_F^\text{HO}$. As before, we have to
pick $\lambda^\text{HO} = 1/2$ and thus $k_F^\text{HO} =
\sqrt{2(N-1)}$ in order to ensure that the energy correction vanishes
for $N \rightarrow 1$. Since we are interested in the energy in the
case of strong interaction, we expand the above general expression to
first order in $1/g$:
\begin{equation}
  \label{eq:astrakharchik-c}
  E^\text{t}_g \approx \frac{N^2}{2}
  - \frac{1}{g} \mathcal{C}
\end{equation}
where the desired contact coefficient is given as
\begin{equation}
    \label{eq:mg-pbc-c-hh}
    \mathcal{C} = \frac{8\sqrt{2}}{3\pi}(N-1)^{3/2} \; .  
\end{equation}
The above LDA expression is directly comparable to our exact
results. Notice that the LDA method introduced ambiguities in the
choice of the Fermi momenta, but that known physics could be used to
restrict the choice and get an unambiguous final
result. Unfortunately, this cannot be done in the calculations based
on Oelkers' free results for a finite $N$ in PBC and HWC.

\subsection{Oelkers (PBC and HWC, finite particle numbers)}
\label{subsec:oelkers}
Before we compare the approximated expression in
Eq.~\eqref{eq:mg-pbc-c-hh} with our exact calculations, we wish to
derive two other approximated expression for the contact
coefficient. These two expressions will rely on the results derived by
Oelkers {\it et al.} in Ref.~\cite{oelkers2006}. In this reference,
the authors calculate the energy of a free strongly interacting system
using PBC and HWC. Contrary to McGuire who assumed large $N$, these
results are valid for finite (but even) $N$, which is what
we ultimately are interested in. Our plan is now to apply the LDA
method in order to get expressions for the trapped systems. We start
by considering the case of PBC.

{\it Periodic boundary condition.} To first order in $1/g$, we find
the free ground state energy to be\footnote{We note a sign error in
  Eq. (14) of Ref.\cite{oelkers2006} as the denominator should read
  $\gamma_i - \gamma_j - i$. Also there appear to be a typo in
  Eq. (16) as the first factor being the energy of $N$ Fermi
  particles, $E_\infty^\text{f}$, is recorded to be
  $\pi^2/(6L^2)(N^3-N)$. But this expression is only valid for $N$
  odd, and since $N$ even is assumed, the first factor should be
  $\pi^2/(6L^2)(N^3 + 2N)$. Finally note the difference of a factor of
  $1/2$ in their definition of the energy.}
\begin{align}
  \label{eq:oe-pbc}
  E^\text{f}_g
  &= \frac{\pi^2}{6 L^2} (N^3 + 2N)
  \left[ 1 - \frac{8}{gL} \right] \nonumber\\
  &= E_\infty^\text{f} - \frac{1}{g} \frac{4\pi^2}{3L^3} (N^3 + 2N).  \; 
\end{align}

We now follow the mapping prescription \eqref{eq:lda-map-ginf} and map
$E_\infty^\text{f} \mapsto E^\text{t}_\infty = N^2/2$ and $k_F^\text{PBC}
\mapsto k_F^\text{HO}$. Using the expressions for 
$k_F^\text{PBC}$ and $k_F^\text{HO}$, we can rewrite the mapping as
\begin{equation}
  \label{eq:map-pbc}
\frac{1}{L} \mapsto
\frac{\sqrt{2(N-\tfrac{3}{2} + \lambda^\text{HO})}}
{\pi(N-2+2\lambda^\text{PBC})} \; ,
\end{equation}
where we have kept the tuning parameters unspecified. Applying the 
above substitution to the result in \eqref{eq:oe-pbc}, we obtain the
energy of the trapped system within LDA:
\begin{equation}
  \label{eq:oe-pbc-c}
  E^\text{t}_g \approx \frac{N^2}{2} - \frac{1}{g} \frac{8\sqrt{2}}{3\pi}
  (N^3 + 2N) \frac{(N-\tfrac{3}{2}+\lambda^\text{HO})^{3/2}}
  {(N-2+2\lambda^\text{PBC})^3} \; .
\end{equation}
In the thermodynamic limit, $N \rightarrow \infty$, the contact
coefficient read off Eq.~\eqref{eq:oe-pbc-c} and the one in
Eq.~\eqref{eq:mg-pbc-c-hh} tend towards the same asymptote given by
$\mathcal{C}_\infty = 8\sqrt{2}/(3\pi) N^{3/2}$, regardless of the
choice of the tuning parameters. This is indeed reassuring, but we are
mostly interested in results for small $N$. Eq.~\eqref{eq:oe-pbc-c} is
derived for even $N$, but let us extend the domain also to odd $N$ in
order to restrict the tuning parameters by requiring the vanishing of
the contact coefficient for $N=1$. Since we also require $\mathcal{C}$
to be non-divergent for finite $N$, we end up with $\lambda^\text{HO}
= 1/2$ and $\lambda^\text{PBC} \in \, ]0;1/2[ \, \cup \,
]1/2;1]$. While this fixes $\lambda^\text{HO}$, there is still
ambiguity in $\lambda^\text{PBC}$. Choosing $\lambda^\text{PBC} = 1$
yields
\begin{equation}
  \label{eq:oe-pbc-c-ho}
  \mathcal{C} =
  \frac{8\sqrt{2}}{3\pi}
  (N^3 + 2N) \frac{(N-1)^{3/2}}{N^3} \; .
\end{equation}

\begin{figure}[tbp!]
  \centering
  \hspace*{-7mm}
  \resizebox{1.15\columnwidth}{!}{
\setlength{\unitlength}{1pt}
\begin{picture}(0,0)
\includegraphics{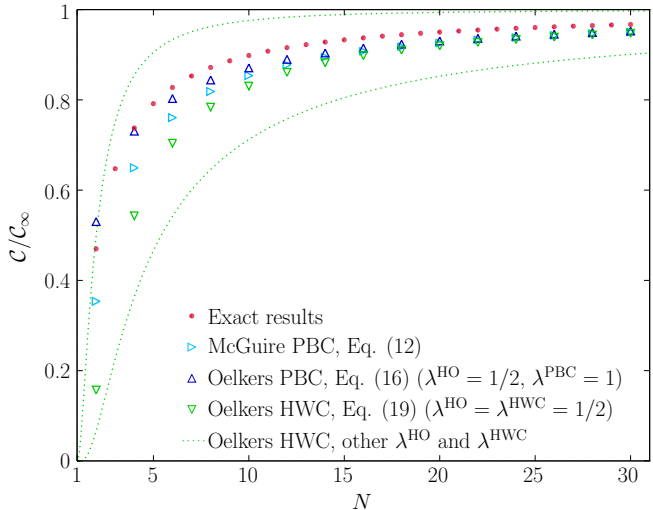}
\end{picture}%
\begin{picture}(576,432)(0,0) \fontsize{17}{0}
  \selectfont\put(74.88,42.5189){\makebox(0,0)[t]{\textcolor[rgb]{0,0,0}{{1}}}}
  \fontsize{17}{0}
  \selectfont\put(134.4,42.5189){\makebox(0,0)[t]{\textcolor[rgb]{0,0,0}{{5}}}}
  \fontsize{17}{0}
  \selectfont\put(208.8,42.5189){\makebox(0,0)[t]{\textcolor[rgb]{0,0,0}{{10}}}}
  \fontsize{17}{0}
  \selectfont\put(283.2,42.5189){\makebox(0,0)[t]{\textcolor[rgb]{0,0,0}{{15}}}}
  \fontsize{17}{0}
  \selectfont\put(357.6,42.5189){\makebox(0,0)[t]{\textcolor[rgb]{0,0,0}{{20}}}}
  \fontsize{17}{0}
  \selectfont\put(432,42.5189){\makebox(0,0)[t]{\textcolor[rgb]{0,0,0}{{25}}}}
  \fontsize{17}{0}
  \selectfont\put(506.4,42.5189){\makebox(0,0)[t]{\textcolor[rgb]{0,0,0}{{30}}}}
  \fontsize{17}{0}
  \selectfont\put(69.8755,47.52){\makebox(0,0)[r]{\textcolor[rgb]{0,0,0}{{0}}}}
  \fontsize{17}{0}
  \selectfont\put(69.8755,117.936){\makebox(0,0)[r]{\textcolor[rgb]{0,0,0}{{0.2}}}}
  \fontsize{17}{0}
  \selectfont\put(69.8755,188.352){\makebox(0,0)[r]{\textcolor[rgb]{0,0,0}{{0.4}}}}
  \fontsize{17}{0}
  \selectfont\put(69.8755,258.768){\makebox(0,0)[r]{\textcolor[rgb]{0,0,0}{{0.6}}}}
  \fontsize{17}{0}
  \selectfont\put(69.8755,329.184){\makebox(0,0)[r]{\textcolor[rgb]{0,0,0}{{0.8}}}}
  \fontsize{17}{0}
  \selectfont\put(69.8755,399.6){\makebox(0,0)[r]{\textcolor[rgb]{0,0,0}{{1}}}}
  \fontsize{17}{0}
  \selectfont\put(298.08,21.5189){\makebox(0,0)[t]{\textcolor[rgb]{0,0,0}{{$N$}}}}
  \fontsize{17}{0}
  \selectfont\put(38.8755,223.56){\rotatebox{90}{\makebox(0,0)[b]{\textcolor[rgb]{0,0,0}{{$\mathcal{C}/\mathcal{C}_\infty$}}}}}
  \fontsize{17}{0}
  \selectfont\put(177.387,160.656){\makebox(0,0)[l]{\textcolor[rgb]{0,0,0}{{Exact
          results}}}} \fontsize{17}{0}
  \selectfont\put(177.387,136.309){\makebox(0,0)[l]{\textcolor[rgb]{0,0,0}{{McGuire
          PBC, Eq. \eqref{eq:mg-pbc-c-hh}}}}} \fontsize{17}{0}
  \selectfont\put(177.387,111.963){\makebox(0,0)[l]{\textcolor[rgb]{0,0,0}{{Oelkers
          PBC, Eq. \eqref{eq:oe-pbc-c-ho} ($\lambda^\text{HO} = 1/2$,
          $\lambda^\text{PBC} = 1$)}}}} \fontsize{17}{0}
  \selectfont\put(177.387,87.6164){\makebox(0,0)[l]{\textcolor[rgb]{0,0,0}{{Oelkers
          HWC, Eq. \eqref{eq:oe-hwc-c-hh} ($\lambda^\text{HO} =
          \lambda^\text{HWC} = 1/2$)}}}} \fontsize{17}{0}
  \selectfont\put(177.387,63.27){\makebox(0,0)[l]{\textcolor[rgb]{0,0,0}{{Oelkers
          HWC, other $\lambda^\text{HO}$ and $\lambda^\text{HWC}$}}}}
\end{picture}
}
\caption{Comparing the approximated contact coefficients based on the
  results of Oelkers {\it et al.} \cite{oelkers2006} and McGuire
  \cite{mcguire1965} with our exact results, scaled with the common
  asymptote for $N\rightarrow\infty$. For Oelkers HWC we also change
  the tuning parameters to see how that influences the result. The
  dotted lines show the choices $\lambda^\text{HO} = 1$ and
  $\lambda^\text{HWC} = 1/4$ (upper curve) and $\lambda^\text{HO} =
  1/2$ and $\lambda^\text{HWC} = 1$ (lower curve), where we for
  clarity have taken $N$ to be a continuous parameter.}
  \label{fig:Contactcoef_plot}
\end{figure}

{\it Hard wall condition.} The ground state energy with HWC is found
in Ref.~\cite{oelkers2006}\footnote{We note a sign error on the second
  term in the front factor in Eq. (19) in Ref.~\cite{oelkers2006}.}:
\begin{equation}
  \label{eq:oe-hwc}
  E^\text{f}_g = \frac{\pi^2}{12 L^2} (2N^3 - 3N^2 + N)
  \left[ 1 - \frac{8}{gL} \cos^2 \left( \frac{\pi}{2N} \right) \right]
  \; .
\end{equation}
This result is consistent with the energy found in
Ref.~\cite{guan2012} in the thermodynamic limit ($N,L\to \infty$ and
$N/L$ constant).  In a completely similar way as before, mapping
$E_\infty^\text{f} \mapsto E^\text{t}_\infty = N^2/2$ and
$k_F^\text{HWC} \mapsto k_F^\text{HO}$ yields the energy of the
trapped system:
\begin{align}
  \label{eq:oe-hwc-c}
  E^\text{t} = &\frac{N^2}{2}
  -\frac{1}{g} \frac{8\sqrt{2}}{3\pi}
  (N^3 - \tfrac{3}{2}N^2 + \tfrac{1}{2}N) \nonumber \\
  & \times \frac{(N-\tfrac{3}{2}+\lambda^\text{HO})^{3/2}}
  {(N-1+\lambda^\text{HWC})^3}
  \cos^2 \left( \frac{\pi}{2N} \right) \; .  
\end{align}
As expected, the contact coefficient goes towards the common asymptote
$\mathcal{C}_\infty$ in the limit $N \rightarrow \infty$. Again we
wish to extend the result to odd $N$, not just even $N$. To ensure
that the contact coefficient is real and well-defined for any integer
$N \geq 1$ and zero at $N=1$, we should pick \emph{either}
$\lambda^\text{HO} = 1/2$ and $\lambda^\text{HWC} = 0$ \emph{or}
$\lambda^\text{HO} \in [1/2;1]$ and $\lambda^\text{HWC} \in \,
]0;1]$. For the choices $\lambda^\text{HO} = \lambda^\text{HWC} =
1/2$, we get
\begin{align}
  \label{eq:oe-hwc-c-hh}
  \mathcal{C} =
  \frac{8\sqrt{2}}{3\pi}
  (N^3 - \tfrac{3}{2}N^2 + \tfrac{1}{2}N)
  \frac{(N-1)^{3/2}}{(N-\frac{1}{2})^3}
  \cos^2 \left( \frac{\pi}{2N} \right) \; .
\end{align}
In deriving Eqs.~\eqref{eq:oe-pbc-c-ho} and \eqref{eq:oe-hwc-c-hh}, we
had to specify values for the tuning parameters, but notice that we
could just as well have chosen other values within certain bounds.

\subsection{Comparing contact coefficients}
We now wish to compare the contact coefficients that we derived in the
previous two subsections. First of all, we want to see, whether the
approximated results compare well with our exact results. Secondly, we
want to investigate how the choice of the tuning parameters interferes
with the result for small values of $N$. On
Figure~\ref{fig:Contactcoef_plot} we have shown our exact results
compared with the three LDA expressions in
Eqs. \eqref{eq:mg-pbc-c-hh}, \eqref{eq:oe-pbc-c-ho} and
\eqref{eq:oe-hwc-c-hh}.

We see that the LDA results tend to undershoot compared to our exact
results. We also wanted to compare HWC results and PBC results,
because our intuition told us that a hard wall box `looked more' like
a harmonic trap than a free periodic potential did. Contrary to our
expectation, the HWC result does not appear to be better than the PBC
results.  On the other hand, the ambiguity in the LDA results
introduced by the choice of the Fermi level and manifested in the
tuning parameters makes it impossible to compare HWC and PBC in a
unique way: We have to choose values for the tuning parameters. On
Figure~\ref{fig:Contactcoef_plot} we also show the HWC result for
other values of the tuning parameters, resulting in drastically
altered contact coefficients. Even for $N \sim 30$, the choice of the
tuning parameters matters, but eventually it will become insignificant
as $N$ becomes larger and larger.

In our opinion this constitute the most important lesson about the LDA
method: For values of $N$ that realistically can be probed
experimentally today or in the coming years, the partially arbitrary
position of the Fermi level has a very large effect on the LDA
predicted contact coefficient. In the light of this conclusion, it
seems like an improbable stroke of luck that McGuire's result found in
the limit $N \rightarrow \infty$ works so well when extrapolated to
the finite $N$ regime.

\begin{figure*}[tbp]
  \centering
  \hspace*{-7mm}
  \resizebox{1.9\columnwidth}{!}{
\setlength{\unitlength}{1pt}
\begin{picture}(0,0)
\includegraphics{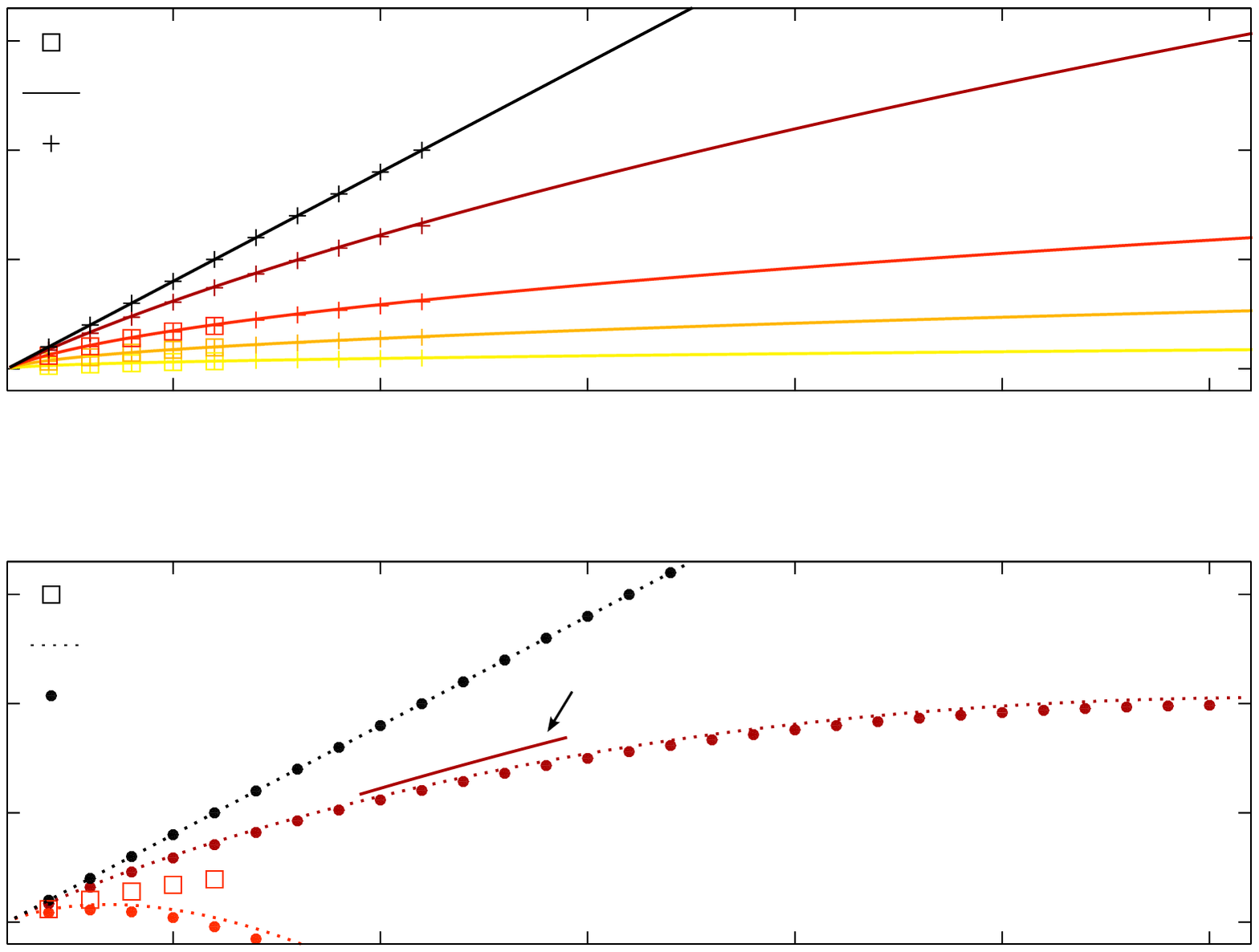}
\end{picture}%
\begin{picture}(576,432)(0,0)
\fontsize{12}{0}
\selectfont\put(74.88,251.32){\makebox(0,0)[t]{\textcolor[rgb]{0,0,0}{{1}}}}
\fontsize{12}{0}
\selectfont\put(134.4,251.32){\makebox(0,0)[t]{\textcolor[rgb]{0,0,0}{{5}}}}
\fontsize{12}{0}
\selectfont\put(208.8,251.32){\makebox(0,0)[t]{\textcolor[rgb]{0,0,0}{{10}}}}
\fontsize{12}{0}
\selectfont\put(283.2,251.32){\makebox(0,0)[t]{\textcolor[rgb]{0,0,0}{{15}}}}
\fontsize{12}{0}
\selectfont\put(357.6,251.32){\makebox(0,0)[t]{\textcolor[rgb]{0,0,0}{{20}}}}
\fontsize{12}{0}
\selectfont\put(432,251.32){\makebox(0,0)[t]{\textcolor[rgb]{0,0,0}{{25}}}}
\fontsize{12}{0}
\selectfont\put(506.4,251.32){\makebox(0,0)[t]{\textcolor[rgb]{0,0,0}{{30}}}}
\fontsize{12}{0}
\selectfont\put(69.8755,264.176){\makebox(0,0)[r]{\textcolor[rgb]{0,0,0}{{0}}}}
\fontsize{12}{0}
\selectfont\put(69.8755,303.402){\makebox(0,0)[r]{\textcolor[rgb]{0,0,0}{{5}}}}
\fontsize{12}{0}
\selectfont\put(69.8755,342.628){\makebox(0,0)[r]{\textcolor[rgb]{0,0,0}{{10}}}}
\fontsize{12}{0}
\selectfont\put(69.8755,381.854){\makebox(0,0)[r]{\textcolor[rgb]{0,0,0}{{15}}}}
\fontsize{12}{0}
\selectfont\put(298.08,236.32){\makebox(0,0)[t]{\textcolor[rgb]{0,0,0}{{$N$}}}}
\fontsize{12}{0}
\selectfont\put(49.8755,324.977){\rotatebox{90}{\makebox(0,0)[b]{\textcolor[rgb]{0,0,0}{{$\Delta E^\text{t}_g$}}}}}
\fontsize{10}{0}
\selectfont\put(246,264.96){\makebox(0,0)[l]{\textcolor[rgb]{0,0,0}{{$g=0.36$}}}}
\fontsize{10}{0}
\selectfont\put(253.44,274.218){\makebox(0,0)[l]{\textcolor[rgb]{0,0,0}{{$g=1.14$}}}}
\fontsize{10}{0}
\selectfont\put(260.88,290.065){\makebox(0,0)[l]{\textcolor[rgb]{0,0,0}{{$g=2.80$}}}}
\fontsize{10}{0}
\selectfont\put(327.84,340.275){\makebox(0,0)[l]{\textcolor[rgb]{0,0,0}{{$g=10$}}}}
\fontsize{10}{0}
\selectfont\put(301.056,377.932){\makebox(0,0)[l]{\textcolor[rgb]{0,0,0}{{$g=10000$}}}}
\fontsize{12}{0}
\selectfont\put(103.718,381.966){\makebox(0,0)[l]{\textcolor[rgb]{0,0,0}{{Experimental data}}}}
\fontsize{12}{0}
\selectfont\put(103.718,363.791){\makebox(0,0)[l]{\textcolor[rgb]{0,0,0}{{Finite $g$, $N \rightarrow \infty$}}}}
\fontsize{12}{0}
\selectfont\put(103.718,345.616){\makebox(0,0)[l]{\textcolor[rgb]{0,0,0}{{Finite $g$, finite $N$}}}}
\fontsize{12}{0}
\selectfont\put(74.88,52.6001){\makebox(0,0)[t]{\textcolor[rgb]{0,0,0}{{1}}}}
\fontsize{12}{0}
\selectfont\put(134.4,52.6001){\makebox(0,0)[t]{\textcolor[rgb]{0,0,0}{{5}}}}
\fontsize{12}{0}
\selectfont\put(208.8,52.6001){\makebox(0,0)[t]{\textcolor[rgb]{0,0,0}{{10}}}}
\fontsize{12}{0}
\selectfont\put(283.2,52.6001){\makebox(0,0)[t]{\textcolor[rgb]{0,0,0}{{15}}}}
\fontsize{12}{0}
\selectfont\put(357.6,52.6001){\makebox(0,0)[t]{\textcolor[rgb]{0,0,0}{{20}}}}
\fontsize{12}{0}
\selectfont\put(432,52.6001){\makebox(0,0)[t]{\textcolor[rgb]{0,0,0}{{25}}}}
\fontsize{12}{0}
\selectfont\put(506.4,52.6001){\makebox(0,0)[t]{\textcolor[rgb]{0,0,0}{{30}}}}
\fontsize{12}{0}
\selectfont\put(69.8755,65.4559){\makebox(0,0)[r]{\textcolor[rgb]{0,0,0}{{0}}}}
\fontsize{12}{0}
\selectfont\put(69.8755,104.682){\makebox(0,0)[r]{\textcolor[rgb]{0,0,0}{{5}}}}
\fontsize{12}{0}
\selectfont\put(69.8755,143.908){\makebox(0,0)[r]{\textcolor[rgb]{0,0,0}{{10}}}}
\fontsize{12}{0}
\selectfont\put(69.8755,183.134){\makebox(0,0)[r]{\textcolor[rgb]{0,0,0}{{15}}}}
\fontsize{12}{0}
\selectfont\put(298.08,37.6001){\makebox(0,0)[t]{\textcolor[rgb]{0,0,0}{{$N$}}}}
\fontsize{12}{0}
\selectfont\put(49.8755,126.257){\rotatebox{90}{\makebox(0,0)[b]{\textcolor[rgb]{0,0,0}{{$\Delta E^\text{t}_g$}}}}}
\fontsize{10}{0}
\selectfont\put(161.184,77.2238){\makebox(0,0)[l]{\textcolor[rgb]{0,0,0}{{$g=2.80$}}}}
\fontsize{10}{0}
\selectfont\put(327.84,123.124){\makebox(0,0)[l]{\textcolor[rgb]{0,0,0}{{$g=10$}}}}
\fontsize{10}{0}
\selectfont\put(301.056,179.212){\makebox(0,0)[l]{\textcolor[rgb]{0,0,0}{{$g=10000$}}}}
\fontsize{12}{0}
\selectfont\put(103.718,183.068){\makebox(0,0)[l]{\textcolor[rgb]{0,0,0}{{Experimental data}}}}
\fontsize{12}{0}
\selectfont\put(103.718,164.893){\makebox(0,0)[l]{\textcolor[rgb]{0,0,0}{{$g \rightarrow \infty$, $N \rightarrow \infty$}}}}
\fontsize{12}{0}
\selectfont\put(103.718,146.718){\makebox(0,0)[l]{\textcolor[rgb]{0,0,0}{{$g \rightarrow \infty$, finite $N$}}}}
\fontsize{12}{0}
\selectfont\put(283,151){\makebox(0,0)[l]{\textcolor[rgb]{0,0,0}{{Finite $g$, $N \rightarrow \infty$}}}}
\end{picture}
}
\caption{Interaction correction to the energy of the non-interacting
  system as function of the total number of particles. The colors
  refer to different values of the interaction strength.  On the top
  figure, we compare experimental data with theories for finite
  $g$. For $N \leq 11$ we have reliable results from a full numerical
  treatment for finite $g$ and finite $N$ ({\bf +}). Notice that the
  agreement with the McGuire formula for finite $g$ and $N \rightarrow
  \infty$ in Eq.~\eqref{eq:astrakharchik} ({\bf ---}) is excellent in
  the few-body regime and is expected to improve for increasing
  $N$. On the bottom figure, we compare experimental data with
  theories assuming $g \rightarrow \infty$. We compare the results for
  finite $N$ using Eq.~\eqref{eq:gs-energy} and the exact geometric
  coefficients in Appendix~\ref{appa}
  (\protect\raisebox{-1.15ex}{\protect\scalebox{3}{$\cdot$}}) and the
  McGuire formula in the limit $N \rightarrow \infty$ in
  Eq.~\eqref{eq:astrakharchik-c} ($\mathbf{\cdots}$). To illustrate
  how the $g \rightarrow \infty$ theories fall off for increasing $N$
  we also show a piece of the finite $g$ McGuire formula for $g=10$ on
  the bottom plot.}
\label{fig:Energy_plot}
\end{figure*}

\section{Comparing models to experiment}
The ground state energy for the impurity system was found
experimentally in Ref.~\cite{wenz2013} for $N=2,\dots,6$ and $g=0.36$,
$1.14$ and $2.80$. Naturally, we want to examine how the theories that
assume large $N$ and/or $g$ compare with each other and the
experimental results.

McGuire assumes that $N \rightarrow \infty$, so we will use
Eqs.~\eqref{eq:astrakharchik} and \eqref{eq:astrakharchik-c} as the
approximated theoretical energies assuming that $g$ is finite and very
large, respectively. We use our exact contact coefficients as a
theoretical prediction assuming finite $N$, but with $g$ being very
large. We also have exact numerical calculations of the energy for
finite $N$ and finite $g$, but this is a very demanding computation,
so we only have reliable results for $N \leq 11$ (using the effective
interaction method recently introduced to address Fermi
\cite{lindgren2014} and Bose systems \cite{amin2015}). These four
theoretical predictions exhaust the four combinations of
finite/infinite $g$ and $N$. We compare them with the experimental
data on on Figure~\ref{fig:Energy_plot}.

Not surprisingly, the most general theory obtained from full numerical
calculations for finite values of $g$ and $N$ is in very good
agreement with the experimental data. But we also see that the theory
assuming finite $g$ and $N \rightarrow \infty$ works equally well for
very small values of $N$. This is indeed an unexpected result,
especially when we recall from the previous section that the exact
position of the Fermi level chosen in the LDA could have a huge impact
on the final result for small $N$.

Let us now examine the theories assuming $g \rightarrow \infty$. We
immediately see that our exact result and the LDA result are in very
good agreement as a direct consequence of the agreement between the
exact contact coefficients and the McGuire result in
Figure~\ref{fig:Contactcoef_plot}. We know that the interaction
correction goes like $\Delta E_g^\text{t} \sim - N^{3/2}/g$ for large
$N$, so for suitable large $N$ and/or small $g$ the theory must break
down (at some point the interaction correction even becomes
negative). This is clearly seen when comparing the top and bottom part
of Figure~\ref{fig:Energy_plot}. It is seen that strongly interacting
theories deviate from the general theories when $N$ increases or $g$
decreases. The interesting question is then for which $N$ and $g$ does
the strongly interaction theory describe nature? The largest
experimental interaction strength probed in Ref.~\cite{wenz2013} is $g
= 2.80$, but as is seen on the bottom part of
Figure~\ref{fig:Energy_plot}, theory and experiment only matches for
$N=2$, so this is clearly not very strong interaction. For $g=10$, we
get a good agreement between the theories assuming finite $g$ and
large $g$ for $N \lesssim 12$, but then the theories split up. We find
it surprising how quickly the strong interaction theory breaks down.

\section{Summary}
We have considered a trapped one-dimensional Fermi system interacting
with a single impurity of the same mass from the point of view of
energetics. We have discussed how the local density approximation
served as a way to map the ground state energy of the free system into
the ground state energy of the trapped system. We also saw that the
method could introduce ambiguous results for finite $N$ that heavily
influenced the predicted energies. This naturally lead to a skepticism
for LDA results for small $N$, which are currently used to describe
experiments\cite{wenz2013}. Despite the erratic nature of the LDA
results, the energy expression based on McGuire's result works
surprisingly well\cite{mcguire1965,astrakharchik2013,wenz2013}.

We also discussed how the strong interaction limit must be approached
with caution as the particle number increase. Comparing theories for
finite interaction and strong interaction, we found that $g = 2.80$
cannot be classified as strong interaction, and for $g = 10$ the
theories start deviating significantly already around $N \sim
12$. This is consistent with the observation that the natural
interaction strength in the harmonic trap is proportional to
$g/\sqrt{N}$ \cite{astrakharchik2013} (see also the earlier 
discussion in Ref.~\cite{astrakharchik2004}).

\paragraph*{Remark.} During preparation of the revised version of 
this manuscript, we became aware of a paper by Deuretzbacher {\it et al.}
\cite{deu2016} which describes a method for computing the 
geometric coefficients and presents results for harmonic traps 
with up to 30 particles using a method that has some common features
with the computational approach used here.

\acknowledgments
The authors wish to thank A.~S.~Dehkharghani and G.~Z\"{u}rn for
supplying us with numerical and experimental data, respectively.
We would also like to thank A.~G. Volosniev for discussions and 
for reading and commenting on the manuscript.
This work is supported in part by the Danish Council for Independent 
Research DFF Natural Sciences and the DFF Sapere Aude program, as
well as by a grant from the Carlsberg Foundation.

\appendix

\section{Geometric coefficients for a harmonic trap}\label{appa}
Here we tabulate the geometric coefficients, $\alpha_i$, in
Eq.~\eqref{eq:heisenberg} in harmonic oscillator units. For
completeness we note that the explicit unit is $(\hbar\omega)^2 b$,
where $b=\sqrt{\hbar/m\omega}$ is the standard oscillator length. To
the right of every indicated value of $N$ we record $\alpha_1, \,
\alpha_2, \dots$ up to $\alpha_{N/2}$ for $N$ even and
$\alpha_{(N-1)/2}$ for $N$ odd, continuing on the next line if
necessary. The remaining geometric coefficients are given by the
symmetry property $\alpha_k = \alpha_{N-k}$ for $k = 1, \dots, N-1$ as
there are $N-1$ coefficients for each $N$. We have calculated all
$N-1$ coefficients for each $N$ and used the symmetry property to
estimate the numerical precision on the coefficients by noting the
digits in $\alpha_k$ in agreement with the digits in
$\alpha_{N-k}$. In this appendix we have at most given 10 significant
figures even though many of our calculations are even more
precise. Generally a higher precision can be achieved for the
price of higher calculation time. The coefficients given here agree
with previous calculations which have been done for 10 particles
or less \cite{volosniev2014,deuret2014,levinsen2014}.

\begin{widetext}
  \begin{table}[htbp]
    \centering
    \begin{tabular}{l@{\hspace{12pt}}l@{\hspace{9pt}}
        l@{\hspace{9pt}}l@{\hspace{9pt}}l@{\hspace{9pt}}
        l@{\hspace{9pt}}l@{\hspace{9pt}}l@{\hspace{9pt}}l}
      $N$ &$\alpha_1 = \alpha_{N-1}$ &$\alpha_2 = \alpha_{N-2}$
          & $\cdots$ &&&&&\\[.2em]
      2  &$\sqrt{2/\pi} =$ &&&&&&&\\
         &0.7978845608 &&&&&&&\\[.2em]
      3  &$3^3/(2^\frac{7}{2} \sqrt{\pi})=$\hspace{-1cm} &&&&&&&\\
         &1.346430196 &&&&&&&\\[.2em]
      4  &1.787645708 &2.346508058 &&&&&&\\[.2em]
      5  &2.166057718 &3.177197531 &&&&&&\\[.2em]
      6  &2.5021784   &3.902098540 &4.357116131 &&&&&\\[.2em]
      7  &2.807397825 &4.552904442 &5.400326410 &&&&&\\[.2em]
      8  &3.088795182 &5.148133894 &6.345017625 &6.738585693 &&&&\\[.2em]
      9  &3.351118130 &5.69971481  &7.214307411 &7.959044430 &&&&\\[.2em]
      10 &3.597730090 &6.215859197 &8.023484036 &9.08789432
         &9.439679835 &&&\\[.2em]
      11 &3.831114914 &6.702502948 &8.78328981  &10.14280881
         &10.81522991 &&&\\[.2em]
      12 &4.053167612 &7.164099681 &9.501616401 &11.13640807
         &12.10509843 &12.42610595 &&\\[.2em]
      13 &4.265372369 &7.604090937 &10.18446319 &12.078086
         &13.32332240 &13.94120413 &&\\[.2em]
      14 &4.468916965 &8.025200651 &10.83651694 &12.97506385
         &14.48046497 &15.37564625 &15.67276485 &\\[.2em]
      15 &4.664769312 &8.429628019 &11.46152119 &13.83303366
         &15.584720   &16.74091572 &17.31574325 &\\[.2em]
      16 &4.853730361 &8.819178529 &12.06252191 &14.65657900
         &16.64259962 &18.04599032 &18.88235477 &19.16023435\\[.2em]
      17 &5.036471720 &9.195355790 &12.64203656 &15.44945457
         &17.65938206 &19.2980535  &20.38221388 &20.92192258\\[.2em]
      18 &5.213563048 &9.559427571 &13.20217385 &16.21478205
         &18.63941758 &20.50296313 &21.82305903 &22.61094361\\
         &22.87289739 &&&&&&&\\[.2em]
      19 &5.385492433 &9.912474388 &13.74472084 &16.95518993
         &19.58634200 &21.66557213 &23.211233   &24.23549760\\
         &24.74584444 &&&&&&&\\[.2em]
      20 &5.552681814 &10.25542595 &14.27120748 &17.67291605
         &20.50323106 &22.78995562 &24.55201753 &25.80233751\\
         &26.54935034 &26.79783773 &&&&&&\\[.2em]
      21 &5.715498860 &10.58908900 &14.78295544 &18.36988415
         &21.39271425 &23.87957538 &25.84986282 &27.31710\\
         &28.29051532 &28.77583530 &&&&&&\\[.2em]
      22 &5.874266240 &10.91416887 &15.28111566 &19.04776222
         &22.25706048 &24.93740208 &27.10856575 &28.7845843\\
         &29.97529569 &30.68723324 &30.92413925 &&&&&\\[.2em]
      23 &6.029268959 &11.23128639 &15.76669765 &19.70800760
         &23.09824363 &25.96600770 &28.33139508 &30.20885539\\
         &31.6087     &32.53825677 &33.00191145 &&&&&\\[.2em]
      24 &6.180760221 &11.54099143 &16.24059255 &20.35190233
         &23.91799356 &26.96763670 &29.52118965 &31.59345070\\
         &33.19521851 &34.33420853 &35.01560955 &35.24241668 &&&&\\[.2em]
      25 &6.328966176 &11.84377367 &16.70359169 &20.98058126
         &24.71783628 &27.94426160 &30.68043376 &32.94144468\\
         &34.73846196 &36.07965    &36.97074938 &37.41540822 &&&&\\[.2em]
      26 &6.474089790 &12.14007146 &17.15640150 &21.59505463
         &25.49912605 &28.89762693 &31.81131634 &34.25553479\\
         &36.24176315 &37.778550   &38.87208994 &39.5265884
         &39.7444875 &&&\\[.2em]
      27 &6.616314036 &12.43027900 &17.59965570 &22.19622631
         &26.26307117 &29.82928451 &32.9157779  &35.5381033\\
         &37.70800940 &39.43437    &40.72377    &41.58088866
         &42.00871 &&&\\[.2em]
      28 &6.755804540 &12.71475230 &18.03392534 &22.78490885
         &27.01075509 &30.7406219  &33.995548   &36.7912671\\
         &39.139756   &41.05017    &42.5294331  &43.58261037
         &44.21317022 &44.42314030 &&\\[.2em]
      29 &6.892711795 &12.99381416 &18.45972710 &23.3618358
         &27.7431536  &31.63288    &35.05217    &38.016916\\
         &40.53927926 &42.6286     &44.29227783 &45.5355
         &46.36230034 &46.775 &&\\[.2em]
      30 &7.027173023 &13.26775837 &18.8775302  &23.9276722
         &28.461149   &32.50720    &36.0870     &39.2167\\
         &41.90861428 &44.172      &46.01515    &47.44300481
         &48.46001639 &49.069      &49.271 &
    \end{tabular}
  \end{table}
\end{widetext}

\end{document}